\documentclass[prb,aps,epsfig,twocolumn,showpacs]{revtex4}
\usepackage{epsfig,amssymb}
\usepackage{psfrag}

\newcommand\beq{\begin{equation}}
\newcommand\eeq{\end{equation}}
\newcommand\bea{\begin{eqnarray}}
\newcommand\eea{\end{eqnarray}}
\newcommand\non{\nonumber}
\newcommand\noi{\noindent}

\newcommand\ra{\rangle}
\newcommand\la{\langle}

\newcommand\bib{\bibitem}

\begin{document}

\textheight=23.8cm

\title{Spin-1 chain with spin-1/2 excitations in the bulk}
\author{Diptiman Sen and Naveen Surendran} 
\affiliation{Centre for High Energy Physics, Indian Institute 
of Science, Bangalore 560012, India}

\date{\today}

\begin{abstract}
We present a spin-1 chain with a Hamiltonian which has three exactly 
solvable ground states. Two of these are fully dimerized, analogous to the 
Majumdar-Ghosh (MG) states of a spin-1/2 chain, while the third is of the 
Affleck-Kennedy-Lieb-Tasaki (AKLT) type. We use variational and numerical 
methods to study the low-energy excitations which interpolate between these 
ground states in different ways. In particular, there is a spin-1/2 excitation
which interpolates between the MG and AKLT ground states; this is the lowest 
excitation of the system and it has a surprisingly small gap. We discuss 
generalizations of our model of spin fractionalization to higher spin chains 
and higher dimensions.
\end{abstract}
\maketitle
\vskip .6 true cm

\section{Introduction}

Quantum spin systems in one dimension have been studied extensively for many 
years. In some seminal papers, Haldane predicted theoretically that integer 
spin chains with nearest neighbor Heisenberg antiferromagnetic interactions 
should have a gap between the ground state and the first excited state 
\cite{haldane}; this was then observed experimentally in a spin-1 system 
\cite{buyers,renard,ma} and confirmed numerically \cite{night,takahashi,white}.
Haldane's analysis used a field theoretic description of the long-distance and
low-energy modes of the spin system \cite{affleck1,fradkin,auerbach,sierra}.
Affleck, Kennedy, Lieb and Tasaki (AKLT) then showed that the ground state of 
the spin-1 chain can be variationally understood as a state in which each 
spin-1 is thought of as a symmetric combination of two spin-1/2's, and the two
spin-1/2's at each site form a singlet with the spin-1/2's of the neighboring
sites \cite{aklt}. The excitations are given by variational states in which 
one of these singlets is replaced by a triplet. It was shown later that the 
AKLT state can be written as a matrix product state \cite{klumper}.

If a spin chain has sufficiently strong next-nearest-neighbor interactions, 
the system is frustrated and its low-energy properties can be quite different 
from those of the unfrustrated system. For instance, the spin-1/2 chain with 
both nearest neighbor ($J_1$) and next-nearest neighbor ($J_2$) 
antiferromagnetic interactions is gapless if $J_2=0$, but is gapped if $J_2 /
J_1 \gtrsim 0.2411$ \cite{okamoto,chitra}. In the latter case, the ground 
state is doubly degenerate as expected by the Lieb-Schultz-Mattis theorem 
\cite{lieb}. In particular, for the Majumdar-Ghosh (MG) model given by $J_2 /
J_1 = 1/2$, the ground states are exactly solvable \cite{majumdar} and consist
of products of nearest-neighbor singlet states as will be described below. The
lowest excited states then consist of spin-1/2's interpolating between the two
ground states \cite{shastry}. Hence the excitations of the MG model have spin 
1/2 in contrast to the excitations of the AKLT model which have spin 1.

The excitations described above exist in the bulk; they contribute to
thermodynamic quantities like the magnetic susceptibility and the specific
heat. In addition to these excitations, a gapped chain with a finite number 
of sites may also have degrees of freedom localized at the edges. For instance,
the AKLT model on an open chain has spin-1/2 degrees of freedom at the edges
\cite{white}; these can be thought of as remnants of the two spin-1/2's of 
which each spin-1 is composed. These edge degrees of freedom have been studied
using field theoretic methods \cite{naveen}. It may be interesting to consider
spin-1 chains which have spin-1/2 excitations in the {\it bulk}, as this 
would provide an example of spin fractionalization. 

Spin fractionalization was first proposed by Faddeev and Takhtajan in the 
spin-1/2 antiferromagnetic chain \cite{faddeev}; the idea is that the 
elementary excitations, called spinons, carry spin-1/2. This was confirmed 
experimentally in a one-dimensional spin-1/2 system $KCuF_3$ \cite{tennant}. 
It was later shown by Anderson and others that spin fractionalization can 
also occur in higher dimensional systems with resonating valence bond ground 
states \cite{anderson,wen}. This idea has been used to understand the 
low-lying excitations in a two-dimensional spin-1/2 system $Cs_2 CuCl_4$ 
\cite{coldea,yunoki}. In contrast to these examples of spin fractionalization
in spin-1/2 systems, we are proposing a model of spin fractionalization in
higher spin systems in this paper.

A spin-1/2 excitation existing in the bulk of a spin-1 chain must clearly 
have two different ground states on its left and right. For instance, the 
ground state on the left could be of the MG type in which each spin-1 forms a 
singlet with one of its neighbors, while the ground state on the right could 
be of the AKLT type. The spin-1/2 excitation can then be thought of as the 
edge degree of freedom of the AKLT part of the chain. To realize this kind of 
an excitation, we require a Hamiltonian for which both MG and AKLT states are 
ground states. We will present such a Hamiltonian in Sec. II; it contains 
interactions involving three neighboring sites. We will present a variational 
estimate of different possible excitations of the model, and will show that 
the spin-1/2 excitation has the lowest variational energy. In Sec. III, we 
will present numerical results for finite chains, with both open and periodic 
boundary conditions. These will confirm that the spin-1/2 excitations indeed 
have the lowest energy; with periodic boundary conditions, such excitations 
must occur in pairs. In Sec. IV, we will discuss how our model can be 
generalized to higher spins and higher dimensions, i.e., how one can 
construct models which have spin $S$ at each site and spin $S'$ excitations 
in the bulk, with $S' < S$. We will make some concluding remarks in Sec. V.

\section{A spin-1 chain} 

\subsection{Hamiltonian and ground states}

We will first present what appears to be the simplest Hamiltonian of an 
infinite spin-1 chain which has exactly three ground states. This Hamiltonian
is motivated by the following arguments. Given three spin-1's ${\vec S}_1, 
{\vec S}_2$ and ${\vec S}_3$, let us define the projection operators $P_S$ 
which projects on to states with total spin $S$, where $S$ can be 0, 1, 2 or 
3. Now consider a three-spin Hamiltonian of the form $h = c_2 P_2 + c_3 P_3$,
where $c_2, c_3 > 0$. The ground states of $h$ are all the states whose total 
spin is equal to 0 or 1; all such states have zero energy. All the excited 
states have strictly positive energies. If we think of each of the spin-1's 
as being a triplet combination of two spin-1/2's, these ground states 
correspond to states in which at least four of the six spin-1/2's form 
singlets amongst each other. The remaining two spin-1/2's can at most form 
a total spin of 1, no matter how they combine with each other. Now, a 
particular Hamiltonian of the above type is $h = S_{tot}^2 (S_{tot}^2 - 2)$,
where $S_{tot}^2 = ({\vec S}_1 + {\vec S}_2 + {\vec S}_3)^2$; this corresponds
to the coefficients $c_2 = 24$ and $c_3 = 120$. This is the simplest 
Hamiltonian with ground state spins being equal to 0 and 1 in the sense that 
it has the lowest possible powers of the spin operators ${\vec S}_i$. 

We now consider a Hamiltonian for the spin-1 chain of the form
\bea H &=& J ~\sum_n ~h_n ~, \non \\
{\rm where} \quad h_n &=& ({\vec S}_{n-1} + {\vec S}_n + {\vec S}_{n+1} )^2 
\non \\
& & \times ~[({\vec S}_{n-1} + {\vec S}_n + {\vec S}_{n+1} )^2 ~-~ 2] ~.
\label {ham} \eea
(We will set the exchange constant $J$ equal to 1). The ground states of this 
Hamiltonian must have at least two singlet bonds within every group of three 
neighboring spins. It is then easy to see that there are three degenerate 
ground states with zero energy of the forms shown in Fig. 1. The analytical 
expressions for these three states are as follows. Let us define the singlet 
combination of two spin-1's at sites $m$ and $n$ as $|S (m,n) \ra = [|1,-1 
\ra_{mn} - |0,0 \ra_{mn} + |-1,1 \ra_{mn}] / \sqrt{3}$, where we have used the
$S^z$ components to label the states. Then the first two ground states of 
(\ref{ham}) are given by tensor products of singlets between nearest neighbors
of the form
\bea |I \ra &=& \prod_{n=-\infty}^\infty ~|S (2n,2n+1) \ra ~, \non \\
{\rm and} \quad |II \ra &=& \prod_{n=-\infty}^\infty ~|S (2n-1,2n) \ra ~. 
\label{ground12} \eea
These are generalizations of the two ground states of the spin-1/2 chain at 
the MG point \cite{majumdar}. 

\begin{figure}[htb]
\begin{center}
\epsfig{figure=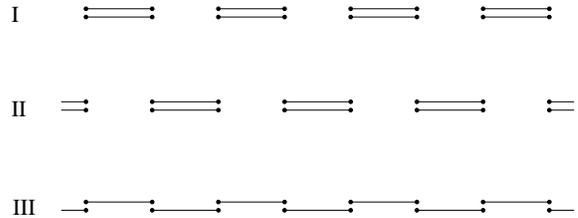,width=8cm}
\end{center}
\caption{The three degenerate ground states. Each solid circle represents a 
spin $1/2$, and the lines denote singlet bonds.}
\end{figure}

The third ground state of Eq. (\ref{ham}) is the AKLT state. This can be 
written as a matrix product state \cite{klumper}. At a site $n$, let us 
define the matrix
\bea M_n &=& \left( \begin{array}{cc} 
\sqrt{1/3} ~|0 \ra_n & \sqrt{2/3} ~|-1 \ra_n \\
- \sqrt{2/3} ~|1 \ra_n & - \sqrt{1/3} ~|0 \ra_n \end{array} \right) ~.
\label{mn} \eea
Then the AKLT state is given by the matrix product
\beq |III \ra ~=~ \prod_{n=-\infty}^\infty ~M_n ~. \label{ground3} \eeq
The matrix in Eq. (\ref{mn}) is motivated as follows \cite{arovas}. For a 
spin-1/2 object, we can use $u=\cos (\theta /2) e^{i\phi /2}$ and $v = \sin 
(\theta /2) e^{-i \phi /2}$ to describe the spin-up and spin-down states 
respectively. The spin operators are given by $S^z = (u \partial_u - v 
\partial_v)/2$, $S^+ = u \partial_v$, and $S^- = v \partial_u$; the total 
spin is $S= (u \partial_u + v \partial_v)/2$. The inner product in the 
$(u,v)$ space is defined by the integration measure $d\Omega = \sin \theta 
d\theta d \phi /(4\pi)$. The correctly normalized spin-1/2 states are given by
$|1/2 \ra = \sqrt{2} u$ and $|-1/2 \ra = \sqrt{2} v$. For a spin-1 object, the
normalized states are given by $|1 \ra = \sqrt{3} u^2$, $|0 \ra = \sqrt{6} 
uv$, and $|-1 \ra = \sqrt{3} v^2$. A singlet formed by spin-1/2's at sites 
$n$ and $n+1$ is given by
\beq u_i v_{i+1} ~-~ v_i u_{i+1} ~=~ \left( \begin{array}{cc} 
u_i & v_i 
\end{array} \right) ~\left( \begin{array}{c} 
v_{i+1} \\
-u_{i+1} \end{array} \right) ~. \eeq
The matrix in Eq. (\ref{mn}) is obtained by combining a column and a row for 
site $n$ as
\beq M_n ~=~ \sqrt{2} ~ \left( \begin{array}{c} 
v_n \\
-u_n \end{array} \right) ~\left( \begin{array}{cc}
u_n & v_n \end{array} \right) ~. \eeq
The normalization of $M_n$ has been chosen so that the norm of the AKLT state 
in Eq. (\ref{ground3}) is given by
\beq Tr \left( \begin{array}{cc}
1/3 & 2/3 \\
2/3 & 1/3 \end{array} \right)^N ~=~ 1 \quad {\rm in ~the ~limit ~N \to 
\infty} ~. \eeq

The three states defined in Eqs. (\ref{ground12}) and (\ref{ground3}) are 
orthonormal for the infinite chain. We do not have an analytical proof that 
these are the only ground states of Eq. (\ref{ham}). However, we will provide 
numerical evidence in Sec. III that there are no other ground states, except 
for some additional degeneracies in open chains due to degrees of freedom at 
the edges.

The structure factor in a ground state is given by
\beq S(q) ~=~ \frac{1}{N} ~\sum_{m,n} ~e^{-iq(m-n)} ~\langle ~{\vec S}_m 
\cdot {\vec S}_n \rangle ~, \eeq
where $N$ is the number of sites in the chain, and we eventually have to take 
the limit $N \to \infty$. In the three ground states given above, we find 
that \cite{arovas}
\bea S^I (q) &=& S^{II} (q) ~=~ 2 ~(1 ~-~ \cos q) ~, \non \\
{\rm and} \quad S^{III} (q) &=& \frac{6 ~(1 ~-~ \cos q)}{5 ~+~ 3 \cos q} ~.
\label{sq} \eea

\subsection{Excited states}

We will now study the excited states using a variational technique
\cite{shastry,caspers,sen}. Given two ground states $A$ and $B$, which could 
be any of the states $I$, $II$ or $III$, one can consider a `domain wall' 
state $|n \ra$ which interpolates between the two at site $n$. We can then 
superpose such states to form momentum eigenstates $|k \ra$ as shown below, 
and obtain a variational estimate of the energy $E_{var} (k) = \la k|H|k 
\ra /\la k|k \ra$. We will now do this for various possible combinations of 
the two ground states $A$ and $B$ on the left and right. There are four 
different cases to consider. In each case, we will form an excited state by 
breaking as few singlet bonds as possible.

\begin{figure}[htb]
\begin{center}
\epsfig{figure=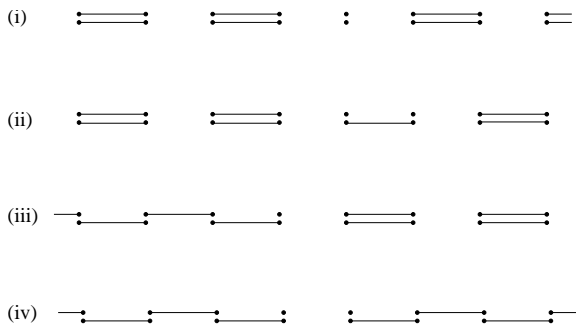,width=8cm}
\end{center}
\caption{Various possible excitations interpolating between different ground 
states. The lines denote singlet bonds, and each isolated circle denotes a 
free spin 1/2.}
\end{figure}

\noi (i) We first consider a state interpolating between ground states $I$ 
on the left and $II$ on the right as shown in Fig. 2 (i). This is given by
\bea |2n (I,II) \ra &=& \prod_{m=-\infty}^n ~|S (2m-2,2m-1) \ra 
\otimes |1 \ra_{2n} \non \\
& & \otimes \prod_{m=n}^\infty ~|S (2m+1,2m+2) \ra ~. \eea
This is a state with $S_{tot}^z = 1$. We then find that
\bea \la 2m (I,II)| 2n (I,II) \ra &=& (1/3)^{|n-m|} ~, \non \\
\la 2m (I,II)| H | 2n (I,II) \ra &=& 40 ~\delta_{m,n} ~. \eea
If we form the momentum eigenstate
\beq |k \ra ~=~ \sum_n ~e^{ik2n} ~|2n \ra ~, \label{mom} \eeq
we find that
\bea \la k|k \ra &=& \frac{2N}{5 ~-~ 3 \cos (2k)} ~, \non \\
{\rm and} \quad \la k|H|k \ra &=& 20 N ~. \label{var1} \eea
{} From Eq. (\ref{var1}), the variational energy is given by
\beq E_{var} (k) ~=~ 10 ~[5 ~-~ 3 \cos (2k)] ~. \eeq
The minimum of this lies at $k=0$, where $E_{var} (0) = 20$.

\noi (ii) Next we consider a state interpolating between ground states $I$
on both the left and the right as shown in Fig. 2 (ii). This is obtained 
by replacing a singlet $|S(2n,2n+1) \ra$ by a triplet. We thus have
\bea |2n (I,I) \ra &=& \prod_{m=-\infty}^n ~|S (2m-2,2m-1) \ra \non \\
& & \otimes ~\frac{1}{\sqrt 2} ~[|1,0 \ra_{2n,2n+1} ~-~ |0,1 
\ra_{2n,2n+1}] \non \\
& & \otimes \prod_{m=n+1}^\infty ~|S (2m,2m+1) \ra ~.
\eea
This is a state with $S_{tot}^z = 1$. We find that
\bea \la 2m (I,I)| 2n (I,I) \ra &=& \delta_{m,n} ~, \non \\
\la 2m (I,I)| H | 2n (I,I) \ra &=& \frac{80}{3} ~\delta_{m,n} ~. \eea
A momentum eigenstate defined as in Eq. (\ref{mom}) satisfies 
\beq \la k|k \ra ~=~ \frac{N}{2} ~, \quad {\rm and} \quad \la k|H|k \ra ~=~ 
\frac{40N}{3} ~. \label{var2} \eeq
Hence the variational energy is 
\beq E_{var} (k) ~=~ 80/3 ~\simeq ~26.67 \eeq 
independent of the value of $k$.

\noi (iii) We now consider a state interpolating between ground states $III$
on the left and $I$ on the right as shown in Fig. 2 (iii). The ground state 
$III$ must end with one singlet bond between the spin-1/2's at site $2n$ and 
$2n+1$, along with a free spin 1/2 at site $2n+1$. We therefore take a state 
which is of the AKLT type from $-\infty$ to site $2n$; this is followed by a 
column multiplied by a free spin 1/2 at the site $2n+1$ of the form
\beq \sqrt{2} ~\left( \begin{array}{c}
v_{2n+1} \\
- u_{2n+1} \end{array} \right) ~u_{2n+1} ~=~
\left( \begin{array}{c}
\sqrt{1/3} ~|0 \ra_{2n+1} \\
- \sqrt{2/3} ~|1 \ra_{2n+1} \end{array} \right) ~. \label{psi3} \eeq
The choice of $u_{2n+1}$, rather than $v_{2n+1}$, as the free spin 1/2 at the 
end of the AKLT region makes this a state with $S_{tot}^z = 1/2$. The free 
spin is then followed on the right by the ground state $I$. The complete 
state is thus given by
\bea |2n (III,I) \ra ~= \prod_{m=-\infty}^{2n} ~M_m
\otimes \left( \begin{array}{c} \sqrt{1/3} ~|0 \ra_{2n+1} \\
- \sqrt{2/3} ~|1 \ra_{2n+1} \end{array} \right) \non \eea
\bea ~~~~~~~~~~~~~~~ \otimes \prod_{m=n+1}^\infty ~|S (2m,2m+1) \ra ~. \eea
We then find that
\bea \la 2m (III,I)| 2n (III,I) \ra &=& (-1/\sqrt{3})^{|n-m|} ~, \non \\
\la 2m (III,I)| H | 2n (III,I) \ra &=& \frac{80}{9} ~\delta_{m,n} ~. \eea
A momentum eigenstate defined as in Eq. (\ref{mom}) satisfies 
\bea \la k|k \ra &=& \frac{N}{2~[2+\sqrt{3} \cos (2k)]}~, \non \\
{\rm and} \quad \la k|H|k \ra &=& \frac{40N}{9} ~. \eea
Hence the variational energy is 
\beq E_{var} (k) ~=~ \frac{80}{9} ~[2 ~+~ \sqrt{3} \cos (2k)] ~. \label{var3}
\eeq
The minimum of this lies at $k=\pi/2$, where $E_{var} (\pi /2) \simeq 2.38$.

\noi (iv) Finally, we consider a state interpolating between ground states
$III$ lying on both the left and the right as shown in Fig. 2 (iv). We 
take the AKLT state on the left to be of the same form as the one discussed 
around Eq. (\ref{psi3}), with $2n$ replaced by $n-1$. The state on the right 
begins with a free spin 1/2 multiplying a row at site $n+1$ of the form
\bea \sqrt{2} ~u_{n+1} ~\left( \begin{array}{cc} 
u_{n+1} & v_{n+1} \end{array} \right) \non \eea
\beq =~ \left( \begin{array}{cc} \sqrt{2/3} ~\la 1|_{n+1} & \sqrt{1/3} ~
\la 0|_{n+1} \end{array} \right) ~. \eeq
This is then followed by a state of the AKLT type from site $n+2$ to $\infty$.
The complete state is thus given by
\bea |n (III,III) \ra ~= \prod_{m=-\infty}^{n-1} ~M_m
\otimes \left( \begin{array}{c} \sqrt{1/3} ~|0 \ra_n \\
- \sqrt{2/3} ~|1 \ra_n \end{array} \right) \non \eea
\beq \otimes \left( \begin{array}{cc} \sqrt{2/3} ~\la 1|_{n+1} & \sqrt{1/3} ~
\la 0|_{n+1} \end{array} \right) \otimes \prod_{m=n+2}^{\infty} ~M_m ~. \eeq
This is a state with $S_{tot}^z = 1$. We then find that
\bea & & \la m (III,III)| n (III,III) \ra \non \\
& & = ~\frac12 ~\delta_{m,n} ~-~ \frac16 ~
(~\delta_{m,n-1} ~+~ \delta_{m,n+1}~) ~, \non \\
& & \la m (III,III)| H | n (III,III) \ra \non \\
& & = ~\frac{320}{27} ~\delta_{m,n} ~-~ \frac{80}{27} ~
(~\delta_{m,n-1} ~+~ \delta_{m,n+1} ~) ~. \eea
A momentum eigenstate defined as 
\beq |k \ra ~=~ \sum_n ~e^{ikn} ~|n \ra \eeq
satisfies
\bea \la k|k \ra &=& \left( \frac12 ~-~ \frac13 \cos k \right) ~N ~, \non \\
{\rm and} \quad \la k|H|k \ra &=& \frac{160N}{27} ~(2 ~-~ \cos k) ~.
\label{var4} \eea
Hence the variational energy is
\beq E_{var} (k) ~=~ \frac{320 ~(2 ~-~ \cos k)}{9 ~(3 ~-~ 2 \cos k)} ~. \eeq
The minimum of this lies at $k=\pi$, where $E_{var} (\pi) \simeq 21.33$.

A comparison between the four kinds of excitations discussed above shows that 
the gaps of excitations of type (i), (ii) and (iv) are given by $20$, $26.67$ 
and $21.33$ respectively, while excitation (iii) has a gap of only $2.38$. We 
note that excitation (i) leaves one triangle unsaturated by two bonds, i.e., 
one group of three neighboring spins has no singlet bonds within themselves; 
this can be seen in Fig. 2. Excitations (ii) and (iv) both leave two triangles
unsaturated by one bond each. Excitation (iii) leaves one triangle unsaturated
by one bond. The minimum energy excitation is of type (iii) which represents 
a `domain wall' interpolating between ground state $I$ (or $II$) and $III$, 
i.e., between ground states of the MG and AKLT types. The gap of $2.38$ for 
this state is much less than the excitation energy of 24 of the three-state 
Hamiltonian $h_n$ appearing in Eq. (\ref{ham}). Further, this state has spin 
1/2 arising from the free spin 1/2 described around Eq. (\ref{psi3}).

We have not tried to improve our variational calculations by considering more
extended states which interpolate between the different ground states. Such 
extended states do not seem to greatly improve the energy estimate 
\cite{caspers}; this is because our ground states have fairly short 
correlation lengths. Further, we will see in Sec. III that the numerical 
result for the lowest excitation gap is not very different from the 
variational estimate obtained above in Eq. (\ref{var3}).

\section{Numerical results}

We will now study the model defined in Eq. (\ref{ham}) using exact 
diagonalization of finite chains, with both open and periodic boundary 
conditions (PBC). We will check whether the three states discussed in Sec. 
II. A are the only ground states, and also what the lowest excitation energy 
is. If the spin-1/2 excitations described in Sec. II. B are indeed the lowest 
energy excitations with a gap $\Delta E$, we would expect the gap for open 
chains to be given by $\Delta E$ while the gap for a chain with PBC should be 
$2 \Delta E$. This is because an open chain may have a single spin-1/2 
excitation with a gap in the bulk, and a gapless spin-1/2 degree of freedom 
localized near one of the edges which compensates for the spin 1/2 in the bulk.
But a chain with PBC can only have excitations in the bulk which have integer 
values of $S_{tot}^z$; hence these must consist of at least two spin-1/2 
excitations.

We have studied chains with $N$ ranging from 5 to 10. In the exact
diagonalization procedure, we used the quantum number $S_{tot}^z$ and symmetry
under parity to reduce the sizes of the Hilbert spaces. For open chains with 
an even number of sites, the degeneracy of ground states is found to be 14. 
This confirms that the three states discussed in Sec II. A exhausts the list 
of all ground states since it can be understood as follows using Fig. 1. There
is one state of type I, 9 states of type II (there are two unpaired spin-1's 
at the edges giving a degeneracy of $3^2$), and 4 states of type III (the two 
dangling spin-1/2's at the edges give a degeneracy of $2^2$). For an open 
chain with an odd number of sites, we find 10 degenerate ground states. This 
can be counted as 3 states each of types I and II arising from an unpaired 
spin-1 at one of the edges, and 4 states of type III due to the two dangling 
spin-1/2's at the edges.

For chains with PBC and an even number of sites, we expect 3 degenerate
ground states corresponding to each of the three types. For an odd number of 
sites, ground states of types I and II are not allowed because they would 
leave one triangle unsaturated; thus we expect a unique ground state of 
type III. These expectations have been confirmed by the numerics.

\vspace*{.3cm}
\begin{figure}[htb]
\begin{center}
\epsfig{figure=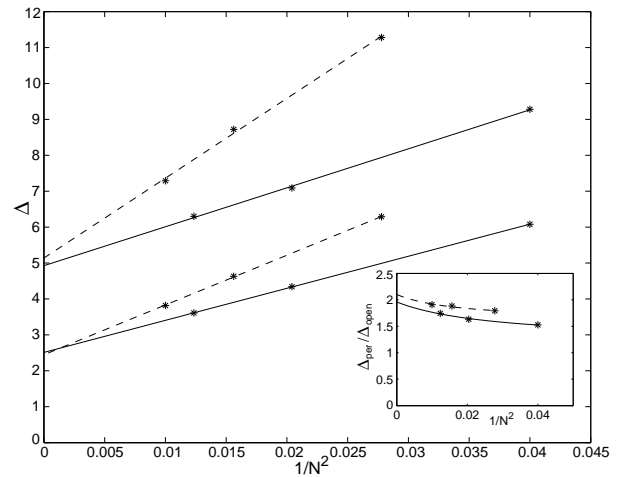,width=8cm}
\end{center}
\caption{Energy gap as a function of $1/N^2$, where $N$ is the chain length.
The upper two lines are for periodic boundary conditions, while the lower two 
lines are for open chains. Dashed and solid lines denote even and odd values 
of $N$ respectively. The inset shows the ratio $\Delta_{per} / \Delta_{open}$
as a function of $1/N^2$.}
\end{figure}

Next we consider the first excited state. Fig. 3 shows the energy gaps as a 
function of the chain length $N$, for chains with PBC (upper two lines) and 
for open chains (lower two lines). Although the results differ significantly 
between even and odd values of $N$, they extrapolate to about the same values 
for $N \to \infty$. We have fitted the gaps to the form $\Delta E (N) = \Delta
E (\infty) + a/N^2$. The reason for this fitting form is that an excited state
with a gap is expected to behave like a particle in a box 
\cite{white,nakamura1}; in a system of length $N$, the leading $N$-dependent
term in the energy of such an object is $1/N^2$. The inset of Fig. 3 shows 
the ratio $\Delta_{per} / \Delta_{open}$ as a function of $1/N^2$ for even 
and odd values of $N$; the lines in the inset are obtained by taking the ratio
of the fitted lines in the main figure.

In Table 1, we summarize the results shown in Fig. 3 by listing the gap for 
various values of $N$ for open ($\Delta_{open}$) and periodic ($\Delta_{per}$)
boundary conditions as well as the ratio $\Delta_{per}/\Delta_{open}$. We see
that the gap for the open chain extrapolates to a value of about $\Delta E = 
2.5$ which is not very different from the value of 2.38 obtained variationally
in Eq. (\ref{var3}). Further, the gap for the chain with PBC extrapolates to 
a value which is about twice that of the open chain gap. This implies, for 
instance, that there is no bound state of two spin-1/2 excitations which has 
an energy which is significantly less than $2 \Delta E$. 

For open chains, we find that the total spin of the lowest excitation is
$S_{tot}=1$ for even $N$ and $S_{tot}=2$ for odd $N$. The latter value can be 
understood as follows: If this excitation is the state (iii) discussed in Sec.
II. B (see Fig. 2 (iii)), which interpolates between AKLT and a fully 
dimerized ground state, then it is possible to have an unpaired spin-1 at the 
edge of the fully dimerized side without costing any energy. This edge spin 
can combine with the spin-1/2 at the edge of the AKLT side and the spin-1/2 
in the bulk to form $S_{tot}=2$.

\begin{table}
\begin{center}
\begin{tabular}{|c|c|c|c|} \hline
$N$ & $\Delta_{open}$ & $\Delta_{per}$ & $\Delta_{per}/\Delta_{open}$ \\ 
\hline
5 & 6.08 & 9.28 & 1.53 \\
6 & 6.29 & 11.28 & 1.79 \\
7 & 4.34 & 7.09 & 1.63 \\
8 & 4.63 & 8.72 & 1.88 \\
9 & 3.61 & 6.30 & 1.75 \\
10 & 3.82 & 7.29 & 1.91 \\
$\infty_{odd}$ & 2.51 & 4.93 & 1.96 \\
$\infty_{even}$ & 2.45 & 5.14 & 2.10 \\ \hline
\end{tabular} 
\end{center}
\caption{Gaps for chains with open and periodic boundary conditions for 
different chain lengths. The last two lines give the gaps extrapolated to 
the thermodynamic limit for an odd and even number of sites respectively.}
\end{table}

\section{Generalizations}

We can construct models involving higher spins or higher dimensions in which 
excitations in the bulk can carry spins which are a fraction of the spin at 
each site. We will discuss some examples below.

\subsection{Higher spin chains}

The idea of a Hamiltonian with multiple ground states in which there are 
varying numbers of singlet bonds between neighboring sites can be generalized
to higher spin chains. Consider a chain of spin-$S$ sites with a Hamiltonian
such that all ground states must have at least $2S$ singlet bonds amongst 
every group of three neighboring sites. In analogy with Eq. (\ref{ham}), we 
can write such a Hamiltonian as $H = \sum h_n$, where $h_n$ is a sum of 
projection operators on to values of total spins ranging from $S+1$ to $3S$ 
for sites $n-1$, $n$ and $n+1$. A state in which there are $p$ singlet bonds 
between sites $2n-1$ and $2n$ and $2S-p$ singlet bonds between sites $2n$ and 
$2n+1$, for every value of $n$, is a ground state of such a Hamiltonian. In 
terms of the variables $u$ and $v$, such a state can be written as 
\cite{nakamura2}
\bea \Psi (p) &=& \prod_{n=-\infty}^\infty ~[(u_{2n-1} v_{2n} - 
v_{2n-1} u_{2n})^p \non \\
& & ~~~~~~~~ (u_{2n} v_{2n+1} - v_{2n} u_{2n+1})^{2S-p}] ~. \label{psip} \eea
Now, each value of $p$ from 0 to $2S$ corresponds to a ground state of the 
Hamiltonian; hence there are $2S+1$ ground states. The case $S=1/2$ 
corresponds to the MG model \cite{majumdar}, while the case $S=1$ corresponds
to the model studied in Secs. II and III. The states in Eq. (\ref{psip})
have appeared in the literature as variational ground states of a dimerized 
spin-$S$ chain, with the integer $p$ changing as the dimerization parameter 
is varied \cite{nakamura2}.

One can now consider excitations which are `domain walls' interpolating
between ground states $\Psi (p_1)$ on the left and $\Psi (p_2)$ on the right,
where, for instance, $p_1 > p_2$. A state of this kind is
\bea & & \Psi_{2n} (p_1,p_2) \non \\ 
&=& \prod_{m=-\infty}^{n-1} ~[(u_{2m-1} v_{2m} - v_{2m-1} u_{2m})^{p_1} \non \\
& & ~~~~~~\times ~(u_{2m} v_{2m+1} - v_{2m} u_{2m+1})^{2S-p_1}] \non \\
& & ~~~~~~\times ~u_{2n-1}^{p_1-p_2} ~\prod_{m=n}^\infty ~[(u_{2m-1} v_{2m} - 
v_{2m-1} u_{2m})^{p_2} \non \\
& & ~~~~~~\times ~(u_{2m} v_{2m+1} - v_{2m} u_{2m+1})^{2S-p_2}] ~. \eea
This state has $S_{tot}^z = (p_1 - p_2)/2$ due to the factor of $u^{p_1-p_2}$
at site $2n-1$. We can now superpose states like this to form a momentum
eigenstate, and calculate its variational energy. A similar procedure can be 
used to construct excited states interpolating between ground states with any
two values of $p_1$ and $p_2$ lying in the range $0 \le p_2 < p_1 \le 2S$. We
thus see that the excited states of this spin-$S$ chain can have any value of 
the spin $(p_1 - p_2)/2$ going from 1/2 to $S$.

\subsection{Higher dimensional models}

One can construct spin models in higher than one dimension in which the 
excited states exhibit spin fractionalization. Two examples are as follows.

\noi (i) Consider a spin-1 model on a square lattice in which the Hamiltonian 
$H$ is a sum over Hamiltonians $H_\square$ of squares for which the ground 
state has at least two singlet bonds in each square \cite{cai}; $H_\square$ 
must be a sum of the projection operators $P_3$ and $P_4$ for the total spin 
of a square. The ground states of $H$ consist of a number of unbroken lines of
singlet bonds such that each square has exactly two such lines running along 
two of its sides. Each line of singlet bonds can either extend all across the 
system or form a closed loop. In the limit of large system size $N$, the 
number of ground states grows as the exponential of $\sqrt{N}$. Hence the 
entropy per site vanishes at zero temperature, even though the number of 
ground states goes to infinity in the thermodynamic limit. Next, we can 
consider excited states in which one of the lines ends at a free spin 1/2 at 
one site; this leaves one square unsaturated. Two such excitations are shown 
in Fig. 4. One can then consider variational states in which the free spin 
1/2 is allowed to move around the lattice in order to reduce its energy.

\begin{figure}[htb]
\vspace*{.8 cm}
\begin{center}
\epsfig{figure=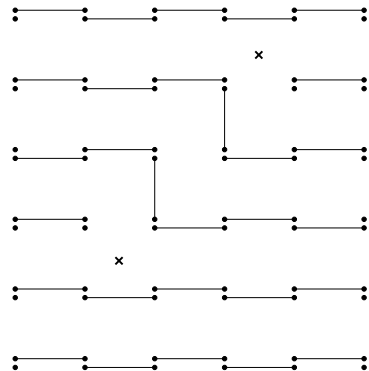,width=5cm}
\caption{Spin-1/2 excitations in a square lattice. For PBC there has to be an 
even number of such excitations. The crosses mark the unsaturated squares 
which have less than two singlet bonds running along their sides.}
\end{center}
\end{figure}

\noi (ii) Next we consider a spin-1 model on a triangular lattice in which the
Hamiltonian $H$ is a sum over Hamiltonians $H_\triangle$ of triangles for 
which the ground state has at least two singlet bonds in each triangle; 
$H_\triangle$ must be the projection operator $P_3$ for the total spin of a 
triangle. The ground states of $H$ consist of unbroken lines of singlet bonds 
such that each triangle has exactly one line running along one of its sides. 
Once again, the number of ground states grows as the exponential of $\sqrt{N}$
for a system with $N$ sites. There are excited states in which one of the 
lines ends at a free spin 1/2 at one site; this leaves one triangle 
unsaturated. The free spin 1/2 can again move around so as to reduce its 
energy.

\section{Conclusions}

We have introduced a Hamiltonian for a spin-1 chain which has three degenerate
ground states, two of the MG type and one of the AKLT type. The lowest energy 
excitation carries spin-1/2 and interpolates between the AKLT state and one of
the MG states; it has a gap $\Delta E \simeq 2.38 J$. In the thermodynamic 
limit $N \to \infty$ and temperatures much lower than $\Delta E /k_B$, the 
system will consist of a dilute gas of the spin-1/2 excitations 
\cite{sen,nakamura1}. Hence a quantity like the magnetic susceptibility will 
go as $\chi \sim \exp (-\beta \Delta E)$ at low temperatures. The spin-1/2 
nature of these excitations can, in principle, be observed in ESR experiments.

Although the model has three ground states, they will not appear with equal 
weights in the limit of very low but non-zero temperature. Since the spin-1/2
excitations interpolate between the AKLT state and either one of the MG 
states, we expect that half the chain will be in the AKLT state, and a quarter
will be in each of the two MG states. This implies that the structure factor 
$S(q)$ at very low temperatures will be given by $[S^I (q) + S^{III} (q)]/2$,
where $S^I (q)$ and $S^{III} (q)$ are given in Eq. (\ref{sq}).

Finally, we have indicated how the spin-1 chain with spin-1/2 excitations can
be generalized to both higher spins and higher dimensions. This provides one 
particular way of realizing the idea of spin fractionalization.

\begin{acknowledgments}
We thank V. Ravi Chandra for discussions, and DST, India for financial support
under the project SP/S2/M-11/2000.
\end{acknowledgments}

\end{document}